# Tuning martensitic transformations via coherent second phases in nanolaminates using free energy landscape engineering

Saaketh Desai[1], Samuel Temple Reeve[2], Karthik Guda Vishnu[1], Alejandro Strachan[1*]

1. School of Materials Engineering and Birck Nanotechnology Center, Purdue University, West Lafayette, IN, USA 47906
2. Materials Science Division, Lawrence Livermore National Laboratory, Livermore, California, 94550, USA

## Abstract

We explore the possibilities and limitations of using a coherent second phase to engineer the thermo-mechanical properties of a martensitic alloy by modifying the underlying free energy landscape that controls the transformation. We use molecular dynamics simulations of a model atomistic system where the properties of a coherent, nanoscale second phase can be varied systematically. With a base martensitic material that undergoes a temperature-induced transformation from a cubic austenite to a monoclinic martensite, the simulations show a significant ability to engineer the transformation temperatures, from a ~50% reduction to a ~200% increase, with 50 at. % of the cubic second phase. We establish correlations between the properties of the second phase and the transformation characteristics and microstructure, via the free energy landscape of the two-phase systems. Coherency stresses have a strong influence on the martensitic variants observed and can even cause the non-martensitic second phase to undergo a transformation. Reducing the stiffness of second phase increases the transformation strain and modifies the martensitic microstructure, increasing the volume fraction of the transformed material. This increase in transformation strain is accompanied by a significant increase in the $A_f$ and thermal hysteresis, while the $M_s$ remains unaltered. Our findings on the tunability of martensitic transformations can be used for informed searches of second phases to achieve desired material properties, such as achieving room temperature, lightweight shape memory alloys.

---

[*] Corresponding author: E-mail: strachan@purdue.edu

# 1. Introduction

Martensitic transformations underlie both shape memory and superelasticity, desirable for a range of applications from connectors and micro-actuators[1] to tires for Mars exploration rovers[2]. The effective design of martensitic materials for these applications hinges on the ability to tune the underlying martensitic transformation for the specific application. For example, low hysteresis is desirable for actuation[1,3], but the opposite is sought for mechanical damping[4]. Such optimizations have traditionally been pursued by modifying the composition of the alloy, using either physics-based approaches[5,6] or, more recently, high-throughput experimental searches which have identified ternary and quaternary alloys with ultra-low thermal hysteresis[7,8]. Machine learning principles coupled with high-throughput density functional theory calculations and experiments have used to discover alloy compositions with ultra-low hysteresis[9]. While these efforts have shown significant success, additional avenues to tune the properties of martensitic materials past composition optimization are desirable, as they can open the design space and potentially result in significantly improved properties. An example of this need is the ß-type family of Mg-Sc martensitic alloys, whose low density (about one third of NiTi based alloys) makes them attractive for aerospace and energy storage applications, yet their low operating temperatures currently make them impractical[10,11]. Specifically, a Mg-20.5 at% Sc alloy showed super elasticity at -150°C while a Mg-19.2 at% Sc alloy showed a thermally induced martensitic transformation starting at -100°C.

The incorporation of coherent second phases has emerged as a novel avenue to tune the thermo-mechanical response of SMAs. This was first demonstrated, via molecular dynamics (MD) simulations, to reduce the hysteresis associated with the martensitic transformation in NiAl alloys by the incorporation of a second phase with desirable characteristics[12]. Recent experiments have shown ultra-low fatigue in NiTi-Cu SMAs via the precipitation of coherent nanoscale $Ti_2Cu$[13], and also have seen favorable changes in transformation characteristics in NiTi-Hf and NiTi-Pt SMAs due to the formation of coherent second phases[14,15] Similarly, nanoscale phase separation via spinodal decomposition in a Ti-Nb gum metal creates a nanoscale composition variation, which in turn results in local confinement of the transformation and super-elasticity over a wide range of temperatures[16]. In addition to second phases obtained through traditional metallurgical processing, epitaxial growth of 5 nm Mg-Nb nanolaminates suppressed the martensitic transformation in Mg, stabilizing the metastable bcc phase at ambient pressure[17]. Similar work has shown the ability to stabilize metastable phases in Cu-Mo thin films[18]. Our previous work with the concept of free energy landscape engineering (FELE) also demonstrated the ability to use coherent second phases to tune transformation characteristics in a controlled manner. Building on Ref. 12, MD simulations have demonstrated that adding a non-martensitic second phase to a martensitic base material, in the form of epitaxial nanolaminates, core-shell nanowires, or nanoprecipitates, can result in reduced thermal hysteresis, tunable transformation temperatures, and even ultra-low stiffness in a fully dense metal or second order martensitic transformations.[19–21] Ab initio simulations have explored strain engineering to increase the martensitic transition temperature in MgSc alloys[22].

While prior work has demonstrated the effect of a specific second phase on transformation characteristics and associated properties[23,24], we lack a general understanding of how the properties of the non-martensitic second phase (relative to the martensitic alloy) map onto the properties of the overall material. Here we use MD simulations to characterize the tunability of martensitic transformation temperatures, thermal hysteresis, and transformation strain in a model system by adding a family of second phases with systematically changing free energy landscapes, with the aim of providing guidelines for choosing precipitates (or other nanostructures) that enable the discovery of novel lightweight SMAs that can operate at room (or elevated) temperature. Our choice of a nanolaminate configuration is partially motivated by the success of strain engineering to enhance semiconductor properties, as exemplified in the increased mobility of strained silicon grown epitaxially on a SiGe layer[38]. Our prior work has documented in detail the microstructure changes for more metallurgically relevant geometries such as precipitates[20].

The remainder of the paper is organized as follows. Section 2 describes the choice of interatomic potential and the procedure used to build the (martensitic) base material, (non-martensitic) second phases, and epitaxial nanolaminates, in addition to providing details for the thermal transformation simulations and free energy landscape calculations. The results of our simulations are described in Sections 3 and 4, focusing separately on the effect of second phase misfit strain and second phase stiffness on transformation characteristics, as compared to the base material, including changes in martensite start ($M_s$) and austenite finish ($A_f$) temperatures, thermal hysteresis, and transformation strain. Finally, we draw conclusions in Section 5.

## 2. Simulation methods

### 2.1 Model martensitic interatomic potential

While metallic alloys, including martensites, are typically described with embedded atom model (EAM) or modified EAM (MEAM) potentials, Elliott et al. developed a generic Morse potential to describe martensitic transformation in binary systems.[25] The potential parameters are a function of a hyperparameter denoted θ, which enables a continuous change in the stability of the martensite and austenite and tuning of the transformation. The potential, accessible through the OpenKIM repository,[26] was developed to describe a Au 47.5 at% Cd SMA (for θ = 400); accurately describing the lattice parameters, thermal expansion coefficients and bulk moduli for the B2 (austenite) and B19 (martensite) phases, in addition to the transformation between the B2 and B19 phases. Since our interest is in a model martensitic material and not in the details of the AuCd system, we will denote the two atom types A and B and treat the potential as one that describes a binary alloy with a high temperature cubic (austenite) phase and, for certain values of the hyperparameter, a transformation to a low temperature monoclinic phase (martensite) and a potential transformation back to the cubic phase. The hyperparameter θ varies the three parameters describing all interactions: cohesive energy, stiffness, and lattice parameter ($D_0$, α, $r_0$) between the different pairs of atom types. For a given θ, varying $r_0$ allows us to simulate a family of second phases with various lattice parameters, but otherwise similar behavior and phase stability. Similarly, slightly reducing the value of the hyperparameter θ results in a second phase with lower stiffness without substantially different phase stability; see Table S1 in the Supplementary Material for more details.

### 2.2 Simulating thermally induced martensitic transformations

We first built a disordered alloy of composition A 47.5 at% B by replicating the B2 unit cell 100 times in the x, y, and z directions, resulting in a system that contains 2,000,000 atoms and with dimensions of 33.5 nm in each direction. Atom types were randomly swapped until the composition of each system was 47.5 at% B. The simulation domain was chosen to be large enough to minimize size effects in the predicted transformation temperatures. The $M_s$ temperature varies strongly as the simulation domain increases from ~16,000 atoms to 1 million atoms where doubling the size to 2 million atoms results in minimal change, see Figure S1 in the Supplementary Information. Finite size are commonly observed in molecular dynamics simulations of phase transitions, as observed before in both solidification[27] and our prior work with NiAl alloys[28]. An important contributor to these size effects in this case is the disordered nature of the alloys as small simulation domains limit the composition heterogeneities present. To simulate thermally-induced martensitic transformations, or lack thereof, in the various systems of interest (base martensitic material, each of the second phases, and the epitaxial nanolaminates) each system was relaxed at 1600 K (above $M_s$ for all systems) for 10 ps under constant stress and temperature (NPT) conditions, allowing all simulation cell angles to evolve independently, using damping constants of 10 fs and 100 fs for the thermostat and barostat respectively. We observed that all stress components were near zero and the potential energy and lattice parameters stabilized after the 10 ps equilibration. Each relaxed structure was then cooled to 200 K at a rate of 5 K/ps under the same NPT conditions through the austenite-martensite transition and subsequently heated back to 2200 K, also at 5 K/ps, through the martensite-austenite transition. All simulations were performed using LAMMPS[29] and the systems visualized using OVITO[30]. Atoms are color coded throughout according to the polyhedral template matching (PTM) analysis[31] with a root mean square deviation cutoff of 0.15, which detects atomic neighborhoods and classifies each atom as BCC (blue, identified as austenite here), HCP (red, martensite), FCC (green, stacking faults), or unidentified (white).

## 2.3 Free energy landscape calculations

The relationship between free energy and lattice parameter of the simulation cell at various temperatures governs the thermodynamics and kinetics of the martensitic transformation[32]. This free energy landscape for each system is calculated by applying a biaxial strain on the austenite phase, in the $[100]_A$ and $[010]_A$ directions (where A refers to austenite). A strain of up to 5% was applied in both tension and compression for the second phases and nanolaminates, while a strain 10% in tension is needed for the base material to cover the full transformation path. The components of the stress and strain tensors are integrated to obtain the free energy along the path:

$$\Delta F = -\int \sigma_{xx} d\epsilon_{xx} + \sigma_{yy} d\epsilon_{yy} + \sigma_{zz} d\epsilon_{zz} + \sigma_{yz} d\epsilon_{yz} + \sigma_{xz} d\epsilon_{xz} + \sigma_{xy} d\epsilon_{xy}$$

where the stress tensor is calculated by LAMMPS, as detailed elsewhere[33] and involves computing the virial, while the strains are computed using the conventional formulae based on changes in the box lengths and angles. We note that the resulting energy landscapes are only approximate representations of the free energy as they depend on the strain rate applied and the path assumed for the transformation (in this case, uniform biaxial deformation). Computing a number of these landscapes and applying Jarzynski's equality[34] can address these limitations of the calculation and relate our non-equilibrium free energy (work or potential of mean force) calculations to the equilibrium free energy landscape. While our approximations do not allow for quantitative predictions of transformation temperatures, they provide useful trends to understand how the properties of the family of second phases vary. All the landscapes shown in this work use a strain rate of $5 \times 10^9$ s$^{-1}$.

## 2.4 Base phases and potential parameters

The hyperparameter θ of the interatomic model allows a description of both martensitic and non-martensitic materials. For θ = 400, resulting in the Morse parameters shown in Table 1, a martensitic transformation occurs with an $M_s$ temperature of 390 K, as shown in Figure 1(a). In some samples we observed a transformation to a different martensite phase (with tetragonal symmetry), see Figure S2 in the Supplementary Material. Since this martensitic phases rarely occurs in the laminate materials studied here, we refer to the monoclinic martensite in the remainder of the paper. For θ = 1000, the resultant parameters describe a non-martensitic alloy that does not transform thermally. The free energy landscapes, shown in Figure 1(b), also describe the martensitic and non-martensitic nature of the materials. At 1000 K, θ = 400 displays a stress-induced transformation, resulting a double-well landscape with equally stable martensite and austenite, while the landscape for the θ = 1000 phase is a single well for the austenite, with no transformation. Figure 1(c) shows the free energy landscapes for θ = 400 across temperature, with the martensite phase increasing in stability as the temperature is decreased.

Table 1: Morse potential parameters obtained for θ = 400 and θ = 1000 in the formulation of Elliott et al.[25]. See Table S1 in the Supplementary Material for more details

| θ | Interaction | $D_0$ | alpha | $r_0$ |
|---|---|---|---|---|
| **400** | A – A | 0.15271 | 1.46152 | 3.15313 |
|  | B – B | 0.48211 | 1.53431 | 3.04440 |
|  | A – B | 0.19979 | 1.76427 | 3.08713 |
| **1000** | A – A | 0.17777 | 1.25703 | 3.33045 |
|  | B – B | 0.43779 | 1.23394 | 3.26694 |
|  | A – B | 0.21675 | 1.61549 | 3.20538 |

To describe epitaxial nanolaminates consisting of both martensitic and non-martensitic phases, we use the random structure generated as described above and add Morse potential parameters for the cross terms, see Figure 1(d). The top half of the cell with the nanolaminate consists of the non-martensitic second phase (atom types C and D) while the bottom half describes the martensitic phase (atom types A and B). Interactions between cross-laminate atom types are then given by mixing rules described by the equations below, similar to mixing rules commonly used in other MD simulations[35]:

$$\alpha_{I-J} = \frac{(\alpha_{I-K} + \alpha_{J-L})}{2}. \qquad r_{I-J} = \frac{(r_{I-K} + r_{J-L})}{2}. \qquad D_{I-J} = \sqrt{D_{I-K} \cdot D_{J-L}}.$$

Where interactions between similar atom types (A and C or B and D) are given by K = I and L = J and interactions between dissimilar atom types (B and C or A and D) are given by (K, L) ∉ (I, J).

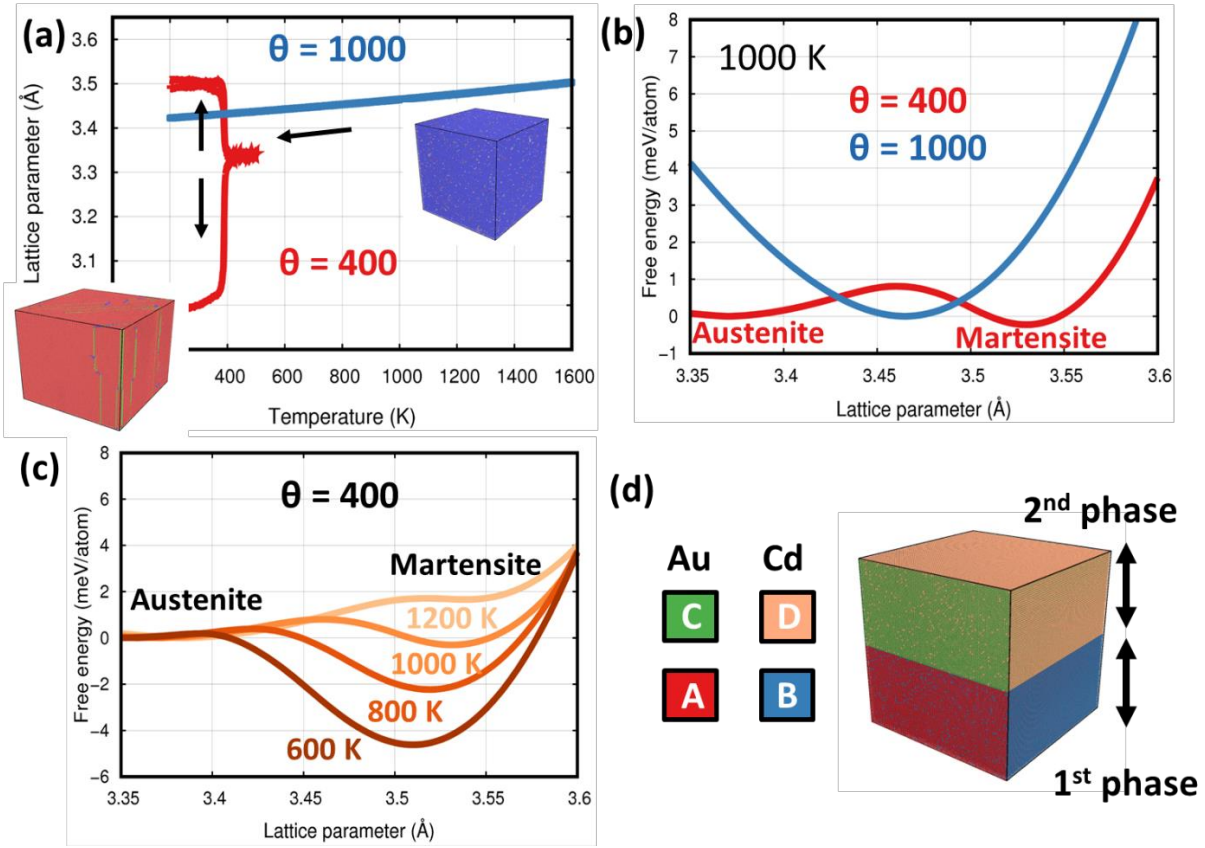

*Figure 1: (a) Cooling simulations showing the martensitic transformation for θ = 400 at ~ 390 K, while θ = 1000 does not transform. The arrows indicate the direction of change in lattice parameter and the inset snapshots show the initial and final structures (austenite and martensite) structures for θ = 400 (b) Free energy landscapes for both phases. The double well structure for θ = 400 shows the stress-induced martensitic transformation, absent for θ = 1000 (c) Free energy landscapes for θ = 400 at various temperatures (d) Initial structure illustrating the 4 atom types used to describe nanolaminates*

# 3. Effect of second phase lattice parameter on transformation characteristics

## 3.1 Second phase lattice parameter between the base material austenite and martensite

### 3.1.1 Effect on transformation temperatures and microstructures

To understand the change in transformation characteristics induced by the lattice parameter of the second phase (relative to the base martensitic material, i.e. misfit strain), we start with six candidate second phases whose lattice parameters span from the austenite to the martensite and epitaxially combine them as nanolaminates with the base martensitic material, described by θ = 400, with 50 at. % of the second phase. The family of second phase materials is described by θ = 1000 in the model Morse potential (resulting in a single cubic phase), with the individual second phases obtained by changing the $r_0$ parameter to obtain the desired range of equilibrium lattice parameters, see Table S2 in the Supplementary Material for the full parameter set. The free energy landscapes of each candidate second phase (denoted P1 to P6) and the base material, at 600 K, are shown in Figure 2(a). The family of second phases ranges from having near zero lattice misfit to the austenite to having near zero misfit to the $[100]_A$ and $[010]_A$ directions of the monoclinic martensite. Note that lattice parameters of the monoclinic martensitic phase and our nanolaminate arrangement allows for near zero in-plane misfit to the martensite despite the difference in symmetry between the two phases.

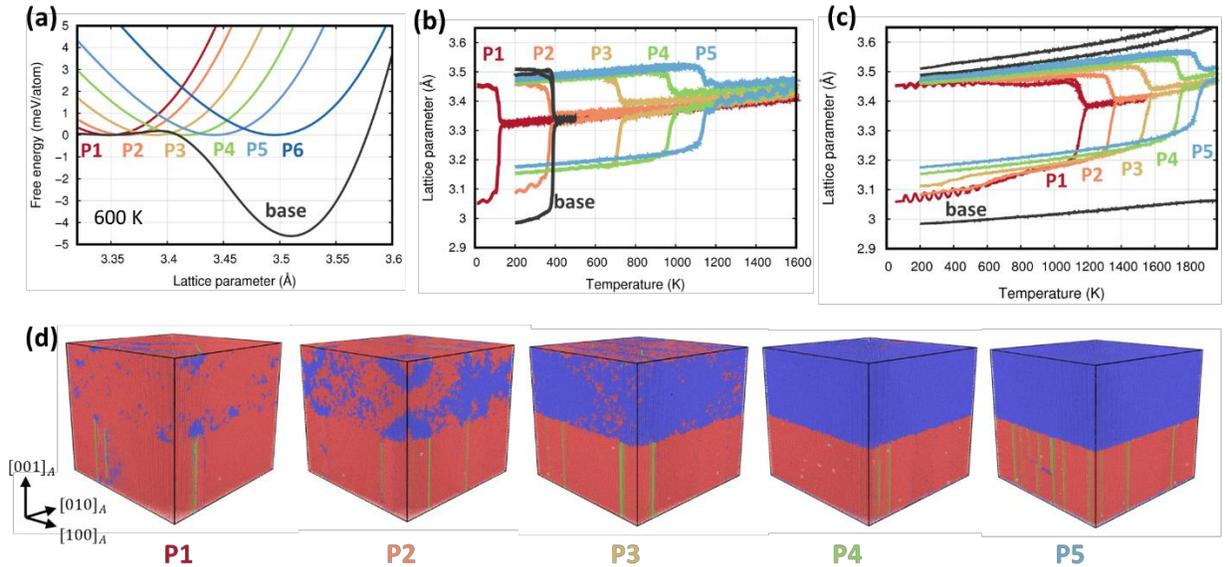

*Figure 2(a): Free energy landscapes of the six candidate phases, denoted P1 to P6, to be epitaxially integrated with the 'base' material, described by θ = 400. Each second phase is non-martensitic, as indicated by the single well energy landscape (b) Cooling simulations showing $M_s$ for nanolaminates with P1 to P5 (c) Heating simulations showing $A_f$ for P1 to P5 nanolaminates*

*and lack thereof for the 'base' material (d) Atomic snapshots (at 200 K), showing the transformed laminates (blue denotes the austenite phase, red martensite, and green defects)*

The cooling simulations, Figure 2(b), indicate a significant ability to modify the transformation temperature via the coherency stresses from the coherent second phase. For this model material, the simulations show that the addition of a second phase can decrease the $M_s$ temperature by up to ~50% or increase it by up to ~200% depending on the lattice mismatch. The $M_s$ temperatures for laminates constructed from candidate phases P1 and P2 (approximately 120 K and 375 K, respectively) are lower than the base material (~ 390 K). On the other hand, laminates constructed from phases P3-P5 (with lower misfit strain with the martensite phase) show $M_s$ temperatures higher than the base material (approximately 730 K, 970 K, and 1130 K respectively). Somewhat surprisingly, see Figure 2(c), all laminates containing any of the P1 to P5 phases result in a martensite to austenite transition upon heating, this not seen in the base material. Thus, adding any of the second phase studied reduces the $A_f$ temperature, even when the austenite phase is stabilized. As expected, this reduction becomes more pronounced as the lattice parameter of the second phase approaches that of the austenite phase. The mechanisms behind this trend are discussed in Section 3.2.

**Effect on microstructure.** In all cases, despite misfit strains approaching 10%, the laminates remain coherent over this wide range of strains due to the non-convex energy landscape of the martensitic phase that results in significantly lower elastic strain than a linear elastic material and the nanoscale dimensions of the laminate periodicity. This is consistent with experimental observations in Fe-Pd magnetic shape memory alloys, where coherent epitaxial growth was achieved for laminates as thick as 50 nm, with the substrate applying strains as large as 8%[36]. We note that the boundary conditions used here make it difficult to lose coherency, where open lateral boundaries would be more appropriate to study coherency limits[37]. Snapshots of these systems at T = 200K, Figure 2(d), show that we form only one martensitic domain whose close packed plane is oriented along $(110)_A$, with stacking faults observed on the $(110)_A$ and $(\bar{1}10)_A$ planes. We also observe that the non-martensitic alloy (top half of the simulation cells) is driven to transform into the martensitic phase (atoms with local martensitic structures are colored red), due to the epitaxial stress caused by the martensitic alloy. The laminate involving the P2 second phase transforms partially and both martensite and austenite phases coexist. For laminates P3-P5, the epitaxial stress from the martensite phase on the second phases is not enough to drive the transformation and the snapshots in Figure 2(d) indicate transformation of only the base martensitic phase, again with a single domain.

Figure 3 and highlights the dependence of $M_s$, $A_f$, and thermal hysteresis on the lattice mismatch of the second phase. For reference, we include the $M_s$ of the base material (dashed red line) and its melting temperature (dashed blue line) since the base material does not have an $A_f$ temperature. The $M_s$ and $A_f$ temperatures increase significantly as the lattice mismatch with the austenite phase increases. A second phase matching the martensite lattice parameter (P6) completely suppresses the martensite to austenite transformation. Under the conditions studied, the austenite to martensite transformation is never completely suppressed, even when the second phase matches the lattice parameter of the austenite phase. We attribute this to the low stiffness of the austenite phase (as compared to the martensite), making it relatively easy to transform to the martensite phase. A larger volume fraction of the second phase or a second phase with higher stiffness would further stabilize the austenite phase and could suppress transformation. Intermediate misfit strains, corresponding to a second phase with a lattice parameter between the austenite and martensitic phases, results in the largest reduction in the activation barrier associated with the transformation and, consequently, lead to the lowest hysteresis. This is consistent with prior results in NiAl alloys[12,20]. We note that our hysteresis values are large compared to experiments; this can be attributed this to the defect-free nature of our initial structures. We have previously observed large hysteresis for defect-free NiAl systems[12,20]. Our results thus indicate a potential avenue to increase the $M_s$ temperature of a martensitic material, as desired for the case of lightweight Mg-Sc shape memory alloys[10,11]. The incorporation of a relatively soft second phase with low misfit strain with the martensite, as demonstrated by second phases P3-P5, could increase the transformation temperature of these alloys to room temperature or above.

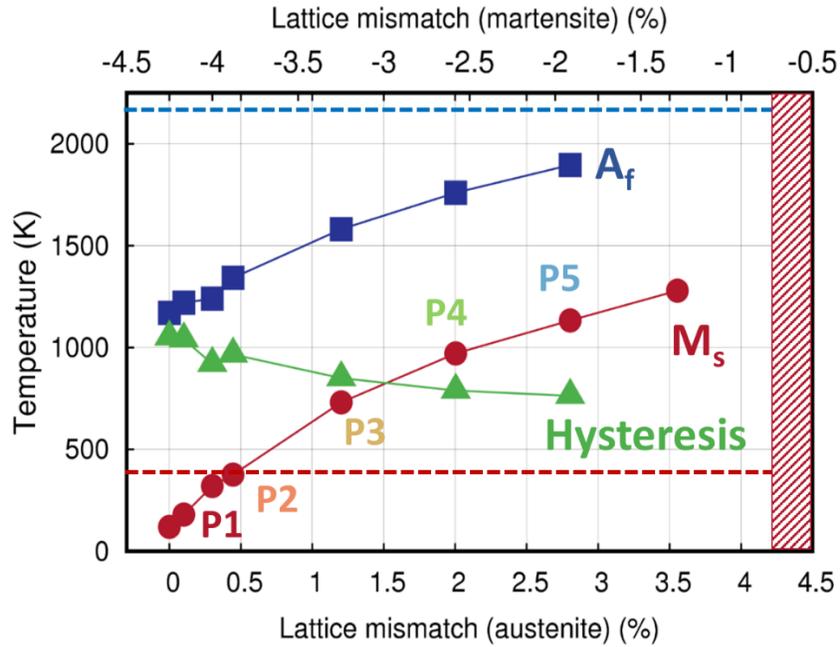

*Figure 3: Trends in $M_s$, $A_f$, and hysteresis as a function of misfit strain (or lattice mismatch) to the austenite and martensite phases of the base material. The red bar indicates a region (phase P6) where the martensite phase is fully stabilized. The dashed red and blue lines represent the $M_s$ and melting temperature (due to lack of $A_f$) of the base material, respectively*

### 3.1.2 Underlying free energy landscapes of the nanolaminates

To understand the trends described in section 3.1.1, we study the free energy landscapes of the family of nanolaminates. We approximate these landscapes by adding the landscapes of the base material and the candidate second phase in equal proportions (since the second phase constitutes 50 at. % of each laminate). Figures 4(a) and 4(b) show the free energy landscapes of the P2 and P4 second phases, respectively, with thin colored lines, the landscape of base alloy in black, and the analytically combined laminate landscapes with thick colored lines. Landscapes are computed at the temperature ($T_0$) where the free energies of the austenite and martensite phase are equal, i.e., the thermodynamic transformation temperature of the laminate (not of the base phase). The features of a landscape that affect the transformation temperature are the energy difference between the martensite and austenite (the thermodynamic driving force) and the barrier for transformation (kinetics).

We first focus on the changes in $M_s$, $A_f$ and hysteresis achieved by adding the second phase, relative to the base material. Figure 4(a) indicates that adding the P2 phase has the effect of stabilizing the austenite with respect to the martensite. The reduction of the driving force to transform to the martensite and slightly increased energy barrier would be expected to result in a lower $M_s$ temperature, which matches the direct cooling simulations. The P1 laminate shows similar behavior. In laminates P3-P5, with P4 as an example in Figure 4(b), the significant reduction in the transformation barrier can be expected to facilitate the martensitic transformation, even with a smaller driving force, increasing $M_s$ as seen in Figure 3. Candidate phase P6 fully stabilizes the martensite, spontaneously transforming to martensite even near the melting temperature, and does not show a martensite to austenite transformation on heating; correspondingly, it has a single well landscape. Regarding the martensite to austenite transition on heating, the reduction in the transformation barrier enables the transformation to austenite that is suppressed in the base material; this is clear in Fig. 4(a) and 4(b). The hysteresis depends on the energy barrier between the austenite and martensite phase at the thermodynamic transformation temperature and Figure 4(c) compares the landscapes for the base material and each nanolaminate. We can confirm that the energy barrier between the austenite and martensite phase is significantly reduced in the laminates as compared with the base alloy, explaining the reduced hysteresis in the thermally induced transformations. Figure 4(d) compares free energy landscapes across phases P1-P6 at a single intermediate temperature, complementing the information presented above and allowing direct comparisons among the second phases themselves. This again confirms that the phase with the lowest thermodynamic transformation temperature (P1) and the highest transformation barrier (requiring large undercooling and overheating) will show the lowest $M_s$ and the lowest $A_f$, and that the $M_s$ and $A_f$ temperatures would increase from P1 to P5, which is what we observe in Figure 3.

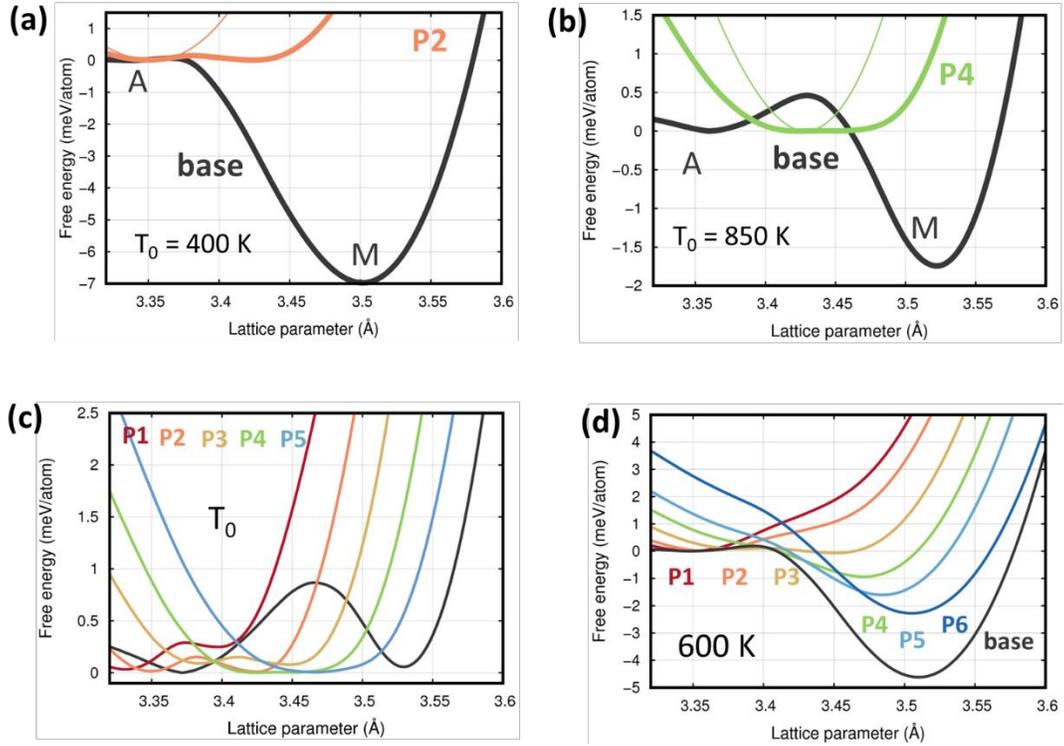

*Figure 4: (a and b) Comparisons of base material landscapes with numerically combined landscapes for laminates P2 and P4, shown as examples. 'A' indicates austenite and 'M' indicates martensite. (c) Free energy landscapes at the thermodynamic transformation temperature $T_0$ (d) Free energy landscapes for each nanolaminate at 600 K. In all landscapes, the horizontal axis is the lattice parameter in the $[100]_A$ and $[010]_A$ directions.*

## 3.2 Second phase lattice parameter beyond the base material austenite

To characterize the limits of FELE in modifying the transformation temperature, we designed a second family of second phases, P1* to P5*, with lattice parameters smaller than those of the austenite phase, see Table S3 in the Supplementary Material for potential parameters. The landscapes for these second phases, in comparison to the base material, are shown in Figure 5(a); direct heating and cooling simulations are shown in Supplementary Figure S3. One could naively expect these second phases to further stabilize the austenite phase relative to the martensite and reduce $M_s$ and $A_f$ further, continuing the trend described in Section 3.1. The cooling simulations, Figure 5(b), show that none of these second phases fully stabilize the austenite. Quite the opposite, phases P4* and P5* stabilize the martensite resulting in $M_s$ temperatures of ~800 K and ~1000 K, comparable with phases P4 and P5. To explain this result, one must consider the difference in symmetry between the phases. Reducing the lattice parameter of the cubic second phase increases the misfit strain with respect to the cubic austenite in both in-plane directions. However, one of the lattice parameters of the monoclinic martensite, is significantly shorter than the other two. Thus, reducing the lattice parameter of the second phase creates an opportunity for a new martensite variant to form where the small lattice parameter accommodates the misfit strain imposed by the lattice mismatch instead of alignment normal to the interface as is the case in the P1 to P5 simulations. The explicit cooling simulations show this, with one martensite variant forming for second phases P1 to P5, while second phases P1* to P5* result in two distinct variants coexisting in elongated domains to accommodate overall strain. Interestingly, we observed phases beyond P5* to fully stabilize the tetragonal martensite phase (see Figure S2 in the Supplementary Material), since the in-plane lattice parameter of the second phase matches the lattice parameter of the tetragonal martensite.

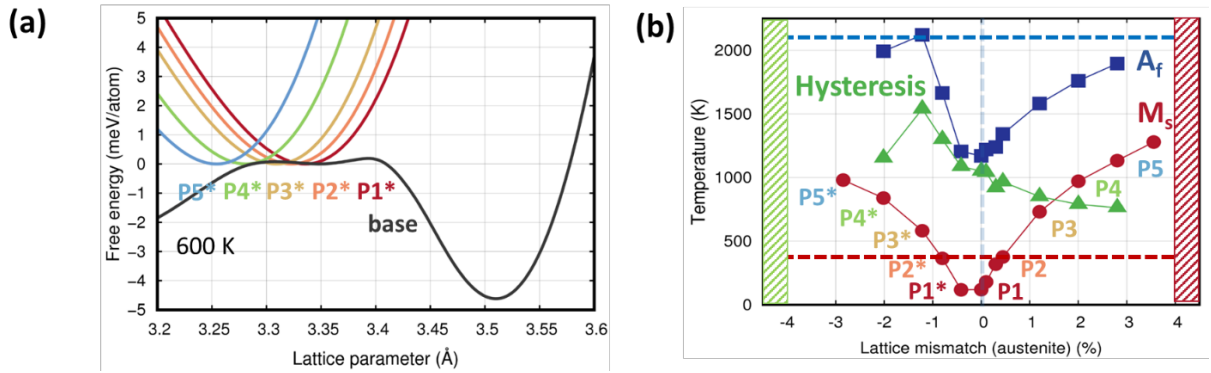

Figure 5: (a) Free energy landscapes of the five candidate phases, denoted P1* to P5*, to be epitaxially integrated over the 'base' material, described by θ = 400 (b) Trends in $M_s$, $A_f$, and hysteresis as a function of misfit strain. The vertical blue line demarcates phases which impose a tensile strain on the base material from phases P1* to P5* which impose a compressive strain. The red bar indicates a region where the monoclinic martensite is fully stable, while the green bar indicates a region where the tetragonal martensite is fully stable. The dashed red and blue lines represent the $M_s$ and melting temperature (due to lack of $A_f$) of the base material, respectively

**Effect on microstructure.** To further understand the relationships between intermediate and negative lattice strains, we estimate the strain energy added to the austenite and the martensite phase, imposed by the lattice mismatch. This is described by the equation below, where $C_{ij}^{\alpha}$ are the elastic constants of the $\alpha$ (austenite or martensite) phase and $\epsilon_{ij}^{\alpha}$ are the strains with respect to that phase.

$$E_{\alpha} = \frac{1}{2} C_{11}^{\alpha} (\epsilon_{11}^{\alpha})^2 + \frac{1}{2} C_{22}^{\alpha} (\epsilon_{22}^{\alpha})^2.$$

The strain energy of the austenite phase increases as the second phase varies from P1 to P6 (see Figure S4), while the strain energy added to the martensite phase decreases, as expected from the landscapes in Figure 2. For second phases P1* to P5*, the rotated martensite variants accommodate the strain such that the strain energy added to the martensite again decreases from P1* to P5*, although to a lesser degree than from P1 to P5. This implies that martensite phase stability with respect to the austenite phase increases from P1* to P5*; this corresponds to increases the $M_s$ and $A_f$ temperatures, as in Section 3.1. We note that this strain energy model only allows us to consider the in-plane lattice mismatch and its effect on the energy difference between the austenite and martensite and does not allow us to comment on the transformation barriers discussed in Section 3.1. We also find that the strain energy added to the austenite and martensite phases by second phases P1*-P5* is comparable in magnitude to phases P1-P5, see Figure S4 in the Supplementary Material. Thus, phases P1* to P5* stabilize the austenite and martensite phases in a similar manner as phases P1 to P5, resulting in similar $M_s$ temperatures.

A consequence of the stabilization of new martensite variants is that phases P1* to P5* show distinct differences in the transformation, particularly in terms of defects generated and the variants of the martensite obtained. Most notably, we observe multiple domains in our microstructures despite the small simulation sizes, where one domain has its close packed plane along $(0\bar{1}\bar{1})_A$ and the other domain has its close packed plane along $(10\bar{1})_A$. The domain wall is oriented along $(\bar{1}\bar{1}0)_A$. Figure 6(a) compares phases P5 and P5*, where the stacking faults (green) are useful in identifying martensite variants of different orientations. The P* family of phases contain combinations of compatible domains, creating a greater number of domains and stacking faults, and therefore retained austenite, upon cooling. Figure 6(b) shows the transition from multi-domain microstructures (P5*) to a single domain (P1*).

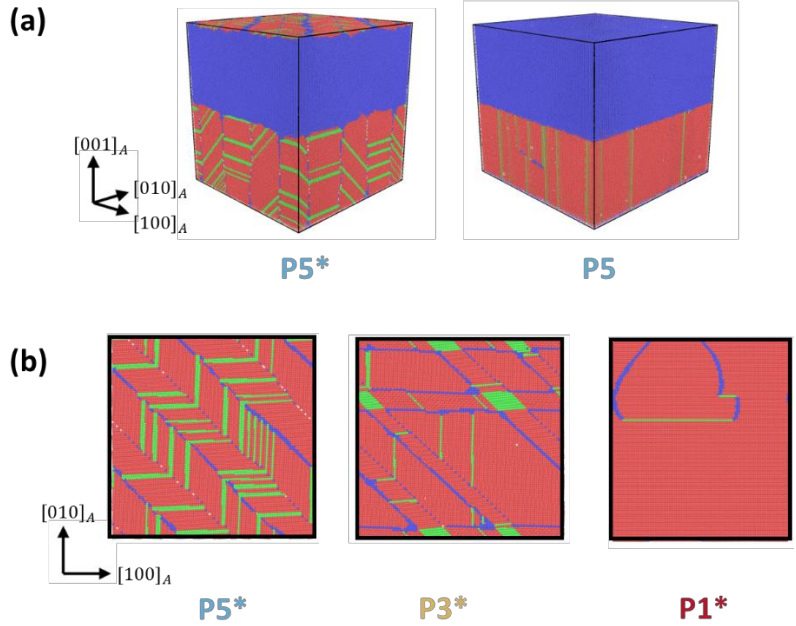

*Figure 6: (a) Atomic snapshots comparing the P5\* laminate to the P5 laminate showing the increased defect formation (b) Slices showing the transition from multi-domain to single domain microstructures from P5\* to P1\**

## 4. Effect of stiffness of the second phase

To understand the effect of second phase stiffness, we now select six additional candidate phases for a third family of phases, $P1^S - P5^S$ this time starting from $\theta = 800$, and again changing $r_0$ to systematically shift the stable lattice parameters, see Table S4 in the Supplementary Material for potential parameters. The free energy landscapes of each of these phases, at 600 K, are shown in Figure 7, clearly much softer than the set of phases from Section 3.1(Fig. 2(a)), indicated by the decreased curvature of the free energy landscape of each of the second phases.

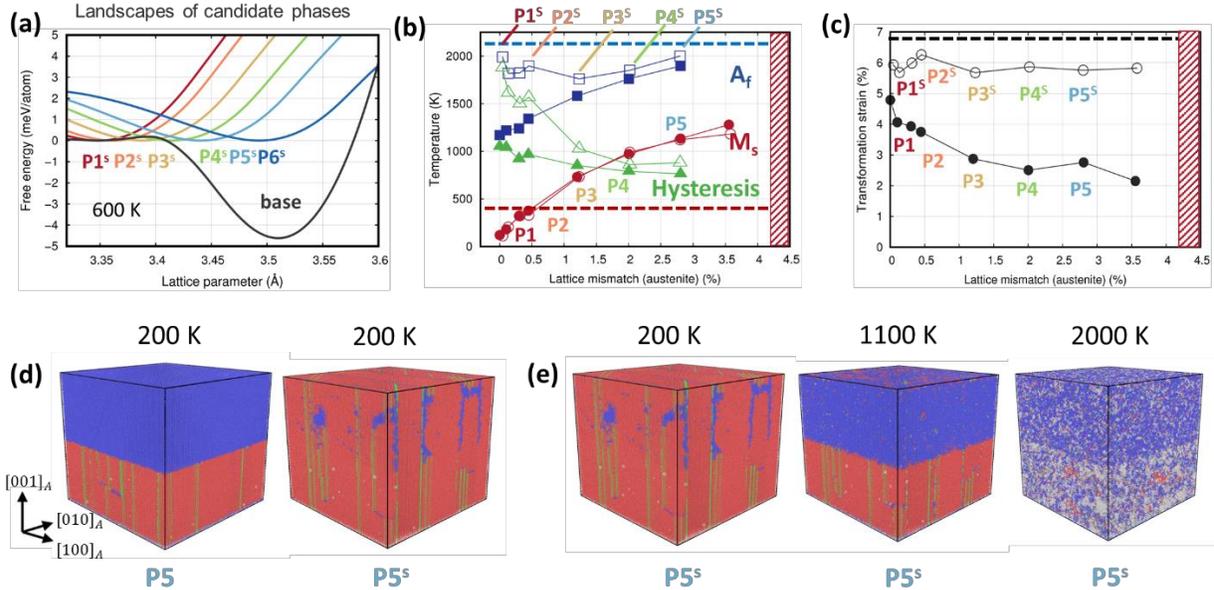

*Figure 7: (a) Free energy landscapes of the six candidate phases, denoted by $P1^S$ to $P6^S$, softer second phases (b) Trends in $M_s$, $A_f$ and hysteresis as a function of lattice mismatch to the austenite and martensite phase. Filled symbols represent P1-P6, open symbols represent $P1^S$-$P5^S$. Refer to Figure 3 for more detail (c) Transformation strain for P1-P5 (closed circles) and $P1^S$-$P5^S$ soft second phases (open circles) as a function of lattice mismatch. Black dashed line represents the transformation strain of the base material. (d) Atomic snapshots showing the transformation for P5 and $P5^S$ (e) Atomic snapshots showing two-step transformation on heating for $P5^S$*

Direct heating and cooling simulations are shown in Supplementary Figure S5. Figure 7(b) again indicates that laminates $P1^S$ and $P2^S$ have an $M_s$ lower than the base material (~110 K and 330 K vs 390 K) while laminates $P3^S$-$P5^S$ have an $M_s$ higher than the base material. The $P1^S$ and $P2^S$ laminates show an $A_f$ of ~2000 K and 1900 K, while the $P3^S$-$5^S$ laminates have an $A_f$ of approximately 1900 K, 1850 K, and 2000 K, respectively. Note again that the base material does not have a well-defined $A_f$. This trend is different from that observed for the stiff set of second phases and will be explored in detail below. The remainder of the trends and observations for this family of second phases follow from the previous sections.

## 4.1 Effect on microstructure

Using a softer set of second phases also allows us to tune the transformation strain, see Figure 7(c). As expected, the softer set of second phases show a greater transformation strain (and transformed volume) as the added second phase transforms from the austenite to the martensite (for all second phases), Figure 7(d). We note that all the candidate second phases belonging to both the soft and stiff set, have transformation strains lower than the base material, both because those added phases are non-transforming on their own, and that all the candidate second phases are stiffer than the base material austenite (in tension). The microstructures observed here again shows a single domain as observed in Section 3.1. In addition, we find laminates P4* and P5* show a distinct two-step transformation while heating from martensite to austenite.

## 4.2 Distinct effects on transformation temperatures

While most trends were observed to be similar between P1-P5 and $P1^S$-$P5^S$, some key distinctions stand out. Most notably, we observe that the $A_f$ temperature for P1 and P2 laminates is now significantly higher than the stiffer second phase laminates (see Fig. 7(b)). To explore this, we directly compare free energy landscapes for the P2-P5 laminates with the $P2^S$-$P5^S$ laminates, see Figure 8. We observe that in all cases, the austenite to martensite transformation barrier is comparable for both the soft and the stiff set of second phases; this matches the fact that the respective $M_s$ temperatures do not differ significantly. However, the martensite to austenite transformation barrier for the P2 and P3 laminates is much higher for the soft set of phases. This similarly matches the $P2^S$ and $P3^S$ laminates' higher $A_f$ compared to P2 and P3 respectively. The differences in landscapes become minor for P4 and P5, as do the differences in $A_f$. Finally, the free energy landscapes also reinforce the fact that using softer second phases results in a greater transformation strain between austenite to martensite.

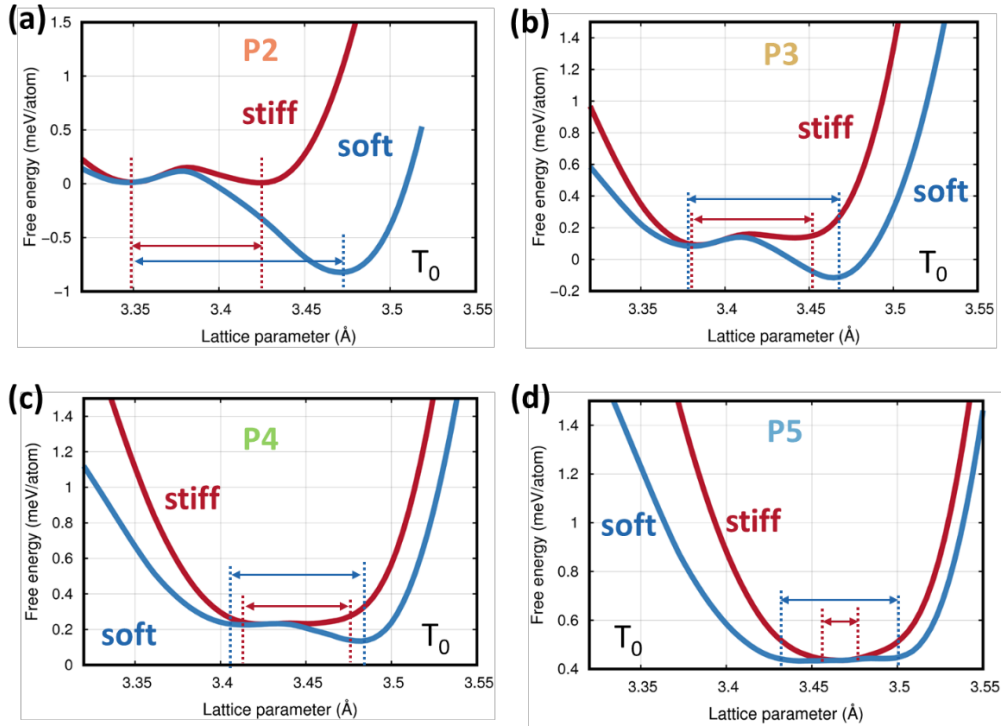

*Figure 8: Direct comparison of numerically combined landscapes for candidate phases (a) P2 and P2$^S$, (b) P3 and P3$^S$, (c) P4 and P4$^S$ and (d) P5 and P5$^S$. The arrows indicate the transformation strain, showing the increase in strain obtained when using softer second phases*

## 5. Conclusions

We systematically investigated the effects of the coherent integration of non-martensitic second phase materials with a base martensitic alloy. We accomplished this by studying a model martensitic system, described by a generic Morse interatomic potential, varying the potential parameters which control lattice parameter and stiffness of the second phases independently. This work is an extension of our previous work in the NiAl family of SMAs, where we demonstrated tunability of $M_s$, $A_f$, and hysteresis for one second phase and multiple volume fractions[12,20].

We find that the $M_s$ temperature can be decreased up to 50% and increased up to 200% (with respect to the base material) by varying the lattice parameter of the second phase, with second phases having a lattice parameter close to the base martensite phase even fully stabilizing the martensite phase. We also observe a reverse transformation (martensite to austenite) in almost all nanolaminates, not seen in the base material, with each of the second phases lowering the martensite to austenite transformation barrier. The $M_s$ and $A_f$ temperatures increase as the lattice mismatch with respect to the austenite phase increases, with a minimum hysteresis observed for a second phase with intermediate lattice mismatch to both the martensite and the austenite phases, again due to a reduction in the transformation barrier. The addition of second phases results in a decrease in the transformation strain for actuation with respect to the base martensitic material due to the mechanical constraints imposed by the non-transforming phase. However, this reduction in transformation strain can be minimized by using a softer second phase, with the caveat of larger hysteresis (as compared to a stiffer non-transforming phase).

This work, therefore, maps the tradeoffs between what is desired: an SMA with large transformation strain, minimal hysteresis, and transformation temperatures near the operating temperature. This can prove to be a guideline for defining and designing second phases that improve SMA characteristics, potentially the operating temperature of lightweight Mg-Sc SMAs[10,11], by incorporating a soft second phase with lattice mismatch to the martensite phase approaching zero. Future work could generalize these trends in a metallurgically relevant precipitate geometry and more specific alloys, providing further guidelines as well as exploring coherency limits.

## Supplementary Material

See supplementary material for complete set of potential parameters and direct heating/cooling simulation results.

## Acknowledgements

This work was supported by the United States Department of Energy Basic Energy Sciences (DoE-BES) program under Program No.DE-FG02-07ER46399 (Program Manager John Vetrano). Computational resources from nanoHUB and Purdue University are gratefully acknowledged.

# [Supplementary Information]

# Tuning martensitic transformations via coherent second phases in nanolaminates using free energy landscape engineering

Saaketh Desai[1], Samuel Temple Reeve[2], Karthik Guda Vishnu[1], Alejandro Strachan[1*]

1. School of Materials Engineering and Birck Nanotechnology Center, Purdue University, West Lafayette, IN, USA 47906
2. Materials Science Division, Lawrence Livermore National Laboratory, Livermore, California, 94550, USA

## Data availability

The datasets generated during and/or analyzed during the current study, including sample input files, potential parameters and LAMMPS data files are available at https://github.rcac.purdue.edu/StrachanGroup/fele_exploration.

## Potential parameters

The full set of potential parameters used for the base material and the stiff and soft set of second phases are given below, see Table S1.

Table S1: Morse potential parameters for the base material and the stiff and soft set of second phases

| Material | Interaction | $D_0$ | alpha | $r_0$ |
|---|---|---|---|---|
| **Base** ($\theta = 400$) | A – A | 0.152716 | 1.46152 | 3.15313 |
|  | B – B | 0.482113 | 1.53431 | 3.04440 |
|  | A – B | 0.199790 | 1.76427 | 3.08713 |
| 2<sup>nd</sup> phases (stiff set, θ = 1000) | A – A | 0.17777 | 1.25703 | 3.19045 – 3.37045 |
| | B – B | 0.437791 | 1.23394 | 3.12694 – 3.30694 |
| | A – B | 0.216752 | 1.61549 | 3.06538 – 3.24538 |
| 2<sup>nd</sup> phases (soft set, θ = 800) | A – A | 0.16684 | 1.34838 | 3.20644 – 3.31644 |
| | B – B | 0.46330 | 1.36743 | 3.11380 – 3.22380 |
| | A – B | 0.20948 | 1.69309 | 3.11024 – 3.22024 |

Table S2: $r_0$ values for each individual phase from the first stiff set of second phases (P1-6), with θ = 1000, $D_0$ and α in Table S1

| $r_0$ | P1 | P1$_1$ | P1$_2$ | P2 | P3 | P4 | P5 | P5$_1$ | P6 |
|---|---|---|---|---|---|---|---|---|---|
| A – A | 3.26045 | 3.26295 | 3.26795 | 3.27045 | 3.29045 | 3.31045 | 3.33045 | 3.35045 | 3.37045 |
| B – B | 3.19694 | 3.19944 | 3.20444 | 3.20694 | 3.22694 | 3.24694 | 3.26694 | 3.28694 | 3.30694 |
| A – B | 3.13538 | 3.13788 | 3.14288 | 3.14538 | 3.16538 | 3.18538 | 3.20538 | 3.22538 | 3.24538 |

Table S3: $r_0$ values for each individual phase from the second stiff set of second phases (P*1-5), with θ = 1000, $D_0$ and α in Table S1

| $r_0$ | P5* | P4* | P3* | P2* | P1* |
|---|---|---|---|---|---|
| A – A | 3.19045 | 3.21045 | 3.23045 | 3.24045 | 3.25045 |
| B – B | 3.12694 | 3.14694 | 3.16694 | 3.17694 | 3.18694 |
| A – B | 3.06538 | 3.08538 | 3.10538 | 3.11538 | 3.12538 |

*Table S4: $r_0$ values for each individual phase from the soft set of second phases (P1-5$^S$), with $\theta$ = 800, $D_0$ and $\alpha$ in Table S1*

| $r_0$ | P1$^S$ | P1$_1$$^S$ | P1$_2$$^S$ | P2$^S$ | P3$^S$ | P4$^S$ | P5$^S$ | P5$_1$$^S$ |
|---|---|---|---|---|---|---|---|---|
| A – A | 3.20644 | 3.20894 | 3.21394 | 3.21644 | 3.23644 | 3.25644 | 3.27644 | 3.29644 |
| B – B | 3.11380 | 3.11630 | 3.12130 | 3.12380 | 3.14380 | 3.16380 | 3.18380 | 3.20380 |
| A – B | 3.11024 | 3.11274 | 3.11774 | 3.12024 | 3.14024 | 3.16024 | 3.18024 | 3.20024 |

The laminate cross-term potential parameters are obtained by the mixing rules provided in Section 2.4 of the main text.

## Size effects on transformation characteristics

The simulation domain was chosen to be large enough to minimize size effects in the predicted transformation temperatures. The $M_s$ temperature varies strongly as the simulation domain increases from ~16,000 atoms to 1 million atoms where doubling the size to 2 million atoms results in minimal change, see Figure S1. We have observed similar size effects in our prior work with NiAl alloys[1,2]. We attribute these size effects to the disordered nature of the alloys as the simulation domain limits the composition heterogeneities present.

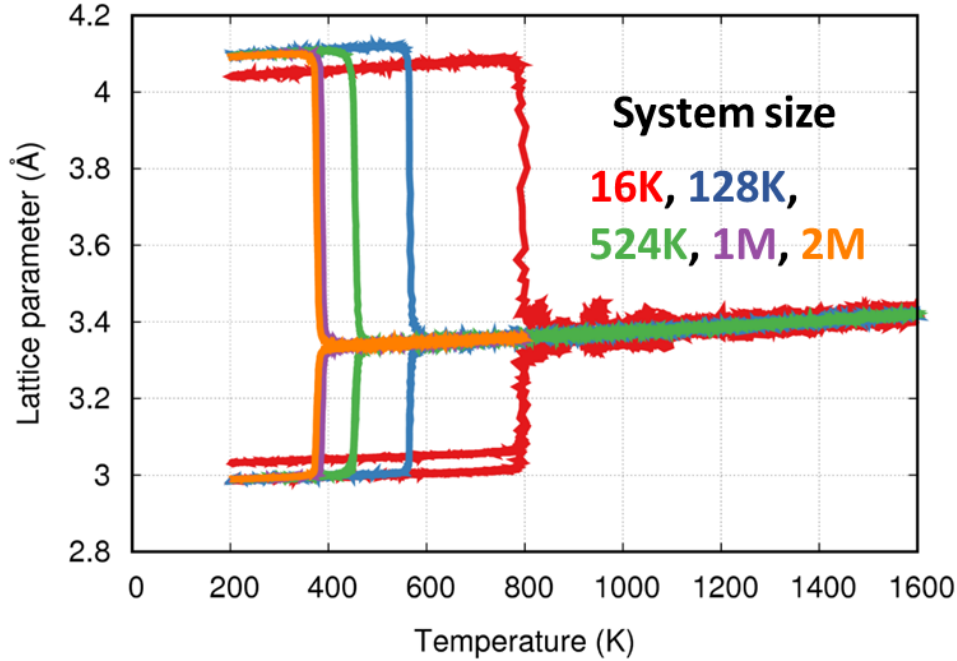

*Figure S1: Size effect on predicted $M_s$ temperature shown via cooling simulations starting from the austenite phase at 1600 K. The labels indicate the number of atoms in the system (~16,000, ~128,000, ~524,000, ~1 million and 2 million atoms respectively). The $M_s$ temperature, detected by the change in lattice parameter varies widely, becomes independent of system size for a system containing greater than 1 million atoms.*

## Other martensitic transformations in the base material

On cooling from the austenite phase, the base material (represented by θ = 400 in the temperature dependent Morse potential parametrization) transforms to a monoclinic as well as a tetragonal martensite, see Figure S2. This is also observed in the free energy landscape, where compression beyond 2-3% results in transformation to the tetragonal martensite phase. This phase is generally suppressed with the cubic second phase, unless that cubic second phase has a significantly shorter lattice parameter that matches the short direction of the martensite.

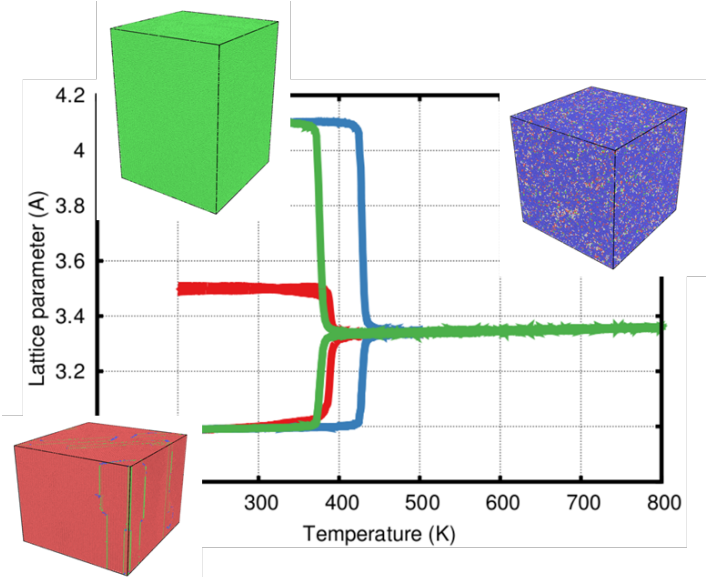

*Figure S2: Cooling simulations of the base material (θ = 400) showing transformations of the austenite phase (blue) to the monoclinic (red) and the tetragonal (green) martensite phases*

## Direct cooling and heating simulations

Figure S3 shows direct heating and cooling simulations that are used to extract the trends detailed in Section 3.2 for phases P1* - P5*.

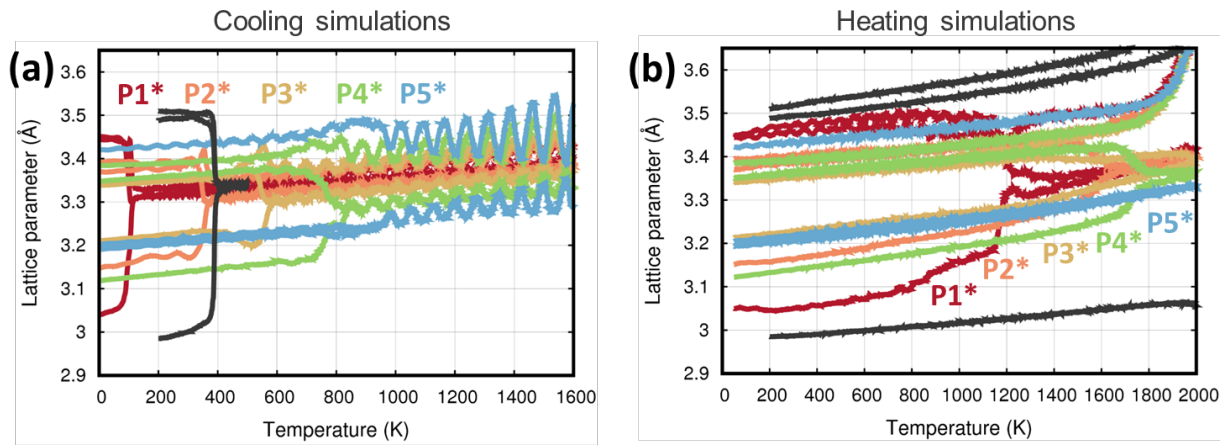

*Figure S3: (a) Cooling and (b) heating simulations for laminates P1* to P5**

# Strain energy added to base material due to second phase

To estimate the strain energy added to the austenite and martensite phases of the base material due to epitaxial integration of a second phase, we propose a simple in-plane strain energy model, see Section 3.2. Figure S4 shows the added strain energy added to both austenite and martensite phases (normalized such that the maximum strain energy added to each phase is 1), highlighting how phases beyond the austenite such as P4* and P5* in particular, reduce the strain energy added to the martensite phase by accommodating multiple domains of two different variants, thus stabilizing the martensite and showing $M_s$ temperatures similar to phases P1-P5, contrary to the naive expectation that P1*-P5* would further stabilize the austenite.

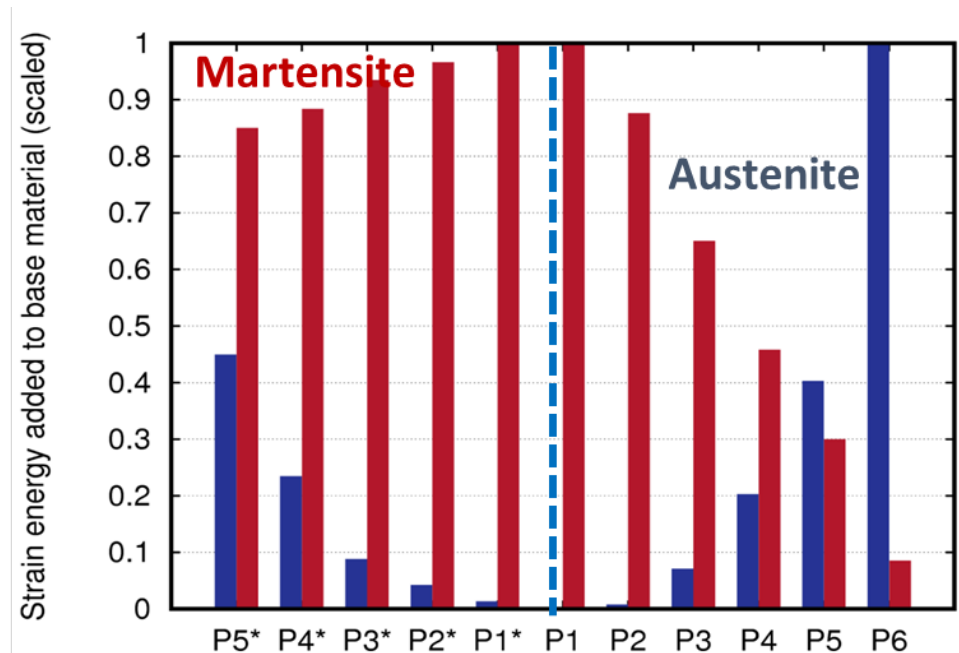

*Figure S4: Scaled strain energy added to the austenite and martensite phases of the base material due to the in-plane and out-of-plane lattice mismatch between the added second phase and the base material. The blue line demarcates phases P1 to P6, which impose a tensile strain (positive in-place lattice mismatch) on the base material, from phases P1* to P5* which impose a compressive strain (negative in-place lattice mismatch) on the base material.*

Figure S5 shows direct heating and cooling simulations that are used to extract the trends detailed in Section 4 for phases $P1^S$ - $P5^S$.

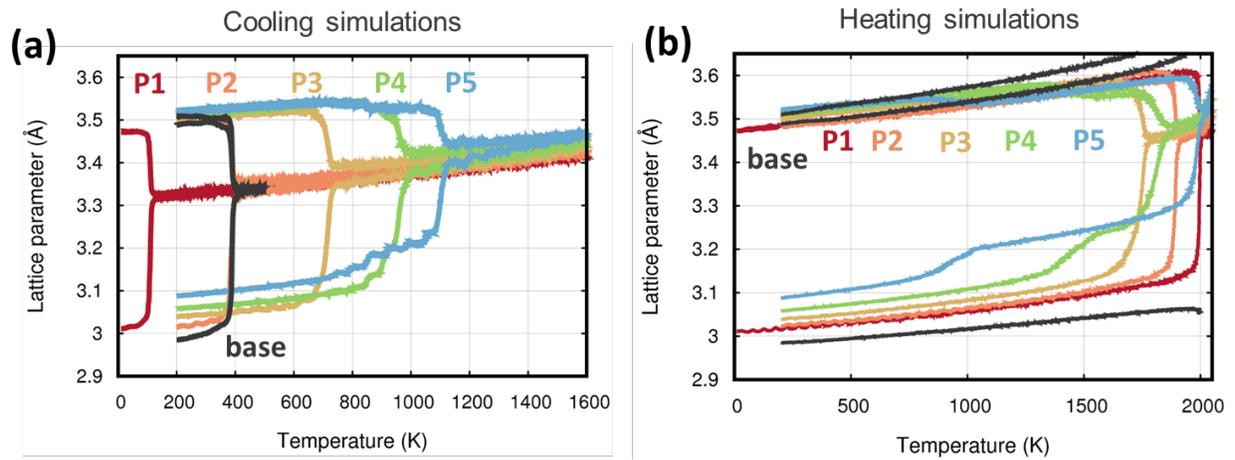

*Figure S5: (a) Cooling and (b) heating simulations for laminates $P1^S$ to $P5^S$*